\documentclass[%
 reprint,
superscriptaddress,
 amsmath,amssymb,
 aps, singlecolumn
]{revtex4-1}

\usepackage{graphicx}
\usepackage{dcolumn}
\usepackage{bm}

\usepackage[latin9]{inputenc}
\usepackage{textcomp}
\usepackage{amssymb}
\usepackage{lmodern}
\usepackage[T1]{fontenc}
\usepackage{hyperref}

\begin{document}

\title{Double inverse nanotapers for efficient light coupling to integrated photonic devices}

\author{Junqiu Liu}
\thanks{J.L. and A.S.R contributed equally to this work}
\affiliation{{\'E}cole Polytechnique F{\'e}d{\'e}rale de Lausanne (EPFL), CH-1015 Lausanne,
Switzerland}

\author{Arslan S. Raja}
\thanks{J.L. and A.S.R contributed equally to this work}
\affiliation{{\'E}cole Polytechnique F{\'e}d{\'e}rale de Lausanne (EPFL), CH-1015 Lausanne,
Switzerland}

\author{Martin H. P. Pfeiffer}
\affiliation{{\'E}cole Polytechnique F{\'e}d{\'e}rale de Lausanne (EPFL), CH-1015 Lausanne,
Switzerland}

\author{Clemens Herkommer}
\affiliation{{\'E}cole Polytechnique F{\'e}d{\'e}rale de Lausanne (EPFL), CH-1015 Lausanne,
Switzerland}

\author{Hairun Guo}
\affiliation{{\'E}cole Polytechnique F{\'e}d{\'e}rale de Lausanne (EPFL), CH-1015 Lausanne,
Switzerland}

\author{Michael Zervas}
\affiliation{{\'E}cole Polytechnique F{\'e}d{\'e}rale de Lausanne (EPFL), CH-1015 Lausanne,
Switzerland}
\affiliation{LIGENTEC SA, CH-1015 Lausanne, Switzerland}

\author{Michael Geiselmann}
\affiliation{{\'E}cole Polytechnique F{\'e}d{\'e}rale de Lausanne (EPFL), CH-1015 Lausanne,
Switzerland}
\affiliation{LIGENTEC SA, CH-1015 Lausanne, Switzerland}

\author{Tobias J. Kippenberg}
\email[]{tobias.kippenberg@epfl.ch}
\affiliation{{\'E}cole Polytechnique F{\'e}d{\'e}rale de Lausanne (EPFL), CH-1015 Lausanne,
Switzerland}

\date{\today}
\begin{abstract}
Efficient light coupling into integrated photonic devices is of key importance to a wide variety of applications. "Inverse nanotapers" are widely used, in which the waveguide width is reduced to match an incident mode. Here, we demonstrate novel, "double inverse" tapers, in which we taper both the waveguide height, as well as the width. We demonstrate that in comparison to regular inverse tapers, the double inverse tapers have excellent polarization-independent coupling. In addition, the optimum coupling is achieved with much larger taper dimension, enabling the use of photolithography instead of electron beam lithography, relevant for applications at near-IR and visible wavelengths. The low coupling loss makes them particularly suitable for nonlinear photonics, e.g. supercontinuum and soliton micro-comb generation.
\end{abstract}
\maketitle


Photonic integrated circuits (PIC) allow for manipulation of light on chip-scale devices and have evolved into a mature technology ("silicon photonics"). Over the past decades, a broad range of applications based on PIC, passive or active, have been demonstrated such as filters \cite{Little:97}, modulators \cite{Reed:05} and lasers \citep{Liang:10}, that are now employed in data-centers. A second wave of interest in integrated photonic devices such as waveguides and resonators, has recently emerged with the advances in their use for nonlinear frequency conversion \cite{Foster:06, Moss:13}. On the one hand, the tight waveguide confinement allows significantly higher effective nonlinearity than that of optical fibers, when combined with materials such as silicon (Si), Si$_3$N$_4$ and AlGaAs that exhibit high material nonlinearity. On the other hand, integrated waveguides enable the engineering of dispersion properties \cite{Foster:06}, in particular the anomalous group velocity dispersion (GVD), a prerequisite for parametric frequency conversion and soliton formation. These advances, along with the fabrication of low loss optical microresonators, have enabled in particular Kerr soliton frequency comb generation in integrated optical microresonators with engineered GVD \cite{Kippenberg:11}, as well as broadband spectra using higher order dispersion engineering via soliton induced Cherenkov radiation \cite{Brasch:15, pfeiffer:17, Li:17}. Moreover, waveguides of engineered dispersion have enabled supercontinuum generation at low pulse energies, i.e. octave spanning combs with pJ pulse energy \cite{Herkommer:17}. These advances signal the potential of integrated nonlinear photonic devices, that have applications such as low noise microwave \cite{Liang:15}, coherent communication \cite{Marin-Palomo:2017}, spectroscopy \cite{Yu:16,Suh:16}, astronomical spectrometer \cite{Obrzud:17,Suh:18} and LIDAR \cite{Trocha:17,Suh:17}.

Central to nonlinear photonic applications is the ability to couple light to photonic chip devices with low loss over a broad optical bandwidth. Widely employed grating couplers \cite{Taillaert:06} are not well suited for this purpose due to their restricted bandwidth. In contrast, inverse nanotapers \citep{Almeida:03}, possessing simultaneously high coupling efficiency, broad operation bandwidth and the use of standard CMOS-compatible fabrication process, are widely used in PIC based chip devices and particularly well suited for nonlinear integrated photonics. An inverse taper works as a mode transformer which adiabatically transforms an incident fiber or free-space optical mode (of several micron mode diameter) to a waveguide mode (of sub-micron mode diameter). The taper mode at the device facet matches the incident fiber or free-space mode, due to the small taper waveguide size and thus the strong evanescent field. 

\begin{figure*}[htb]
\centering
\includegraphics[clip,scale=0.99]{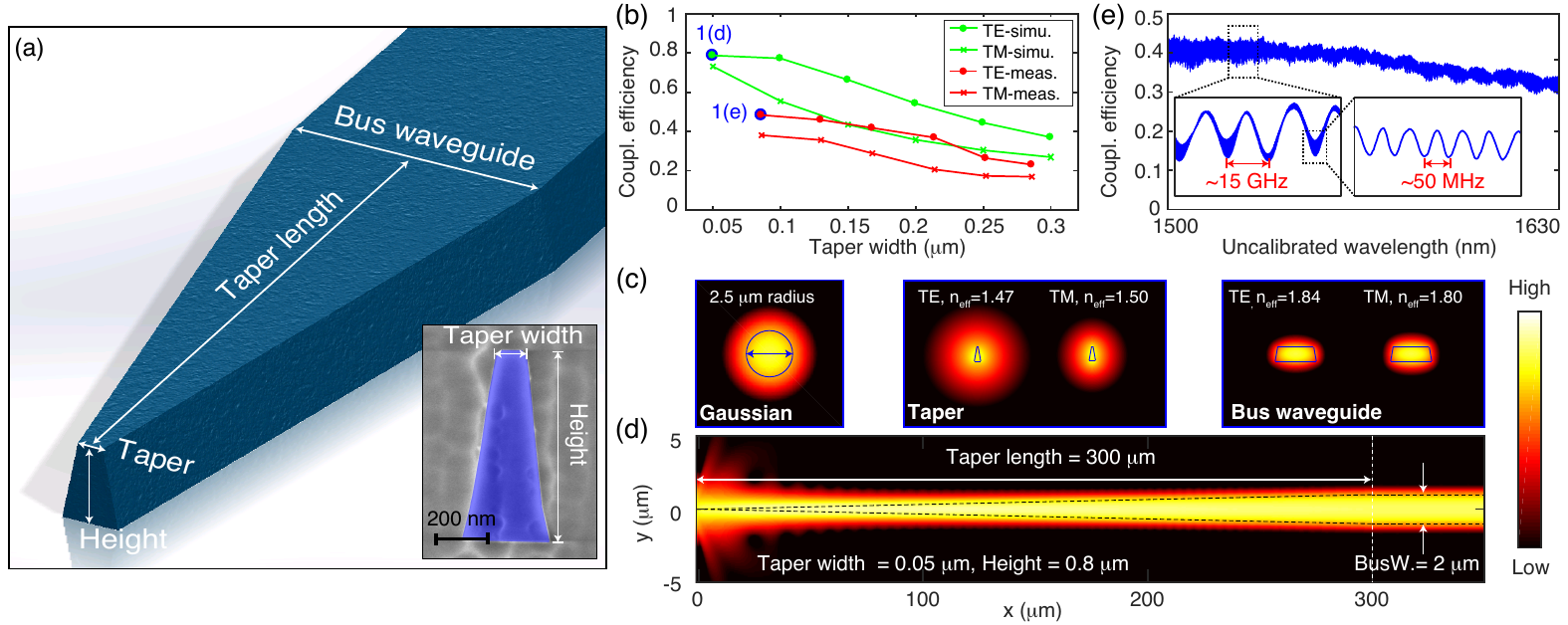}
\caption{Schematic, simulation and characterization of the regular 1D inverse tapers. (a) Schematic of a 1D inverse taper of increasing width and constant height over the taper length. Inset: SEM image of the taper cross-section (Si$_3$N$_4$ is blue shaded), 80 nm in width and 820 nm in height, at the chip facet, buried in SiO$_2$ cladding. The taper width is defined as the trapezoid's top side width. (b) Simulated coupling efficiency (including two chip facets) versus different taper widths for both the TE and TM polarizations (green), in comparison with the experimentally measured data (red). Blue shaded data points are studied with more details in (d), (e). (c) Simulated mode profiles of the incident Gaussian mode, 0.1-$\mathrm{\mu}$m-width taper's TE and TM modes, and 2-$\mathrm{\mu}$m-width bus waveguide's TE and TM fundamental modes, to illustrate the taper's working mechanism as a mode transformer, to bridge the incident Gaussian mode and the bus waveguide mode. $\mathrm n_{\mathrm e \mathrm f \mathrm f}$: effective refractive index. (d) Simulated mode coupling profile in the case of $\sim$80$\%$ coupling efficiency marked in (b). (e) Characterized coupling efficiency from 1500 to 1630 nm, of the taper marked in (b). Two Fabry-Perot interference patterns are observed. The $\mathrm \sim$15 GHz one is due to the reflection between two chip facets (5 mm cavity length), and the $\mathrm \sim$50 MHz one is likely due to the reflection between the input chip facet and the laser (2 m cavity length).}
\label{Fig:figure1}
\end{figure*}

\begin{figure*}[htb]
\centering
\includegraphics[clip,scale=0.99]{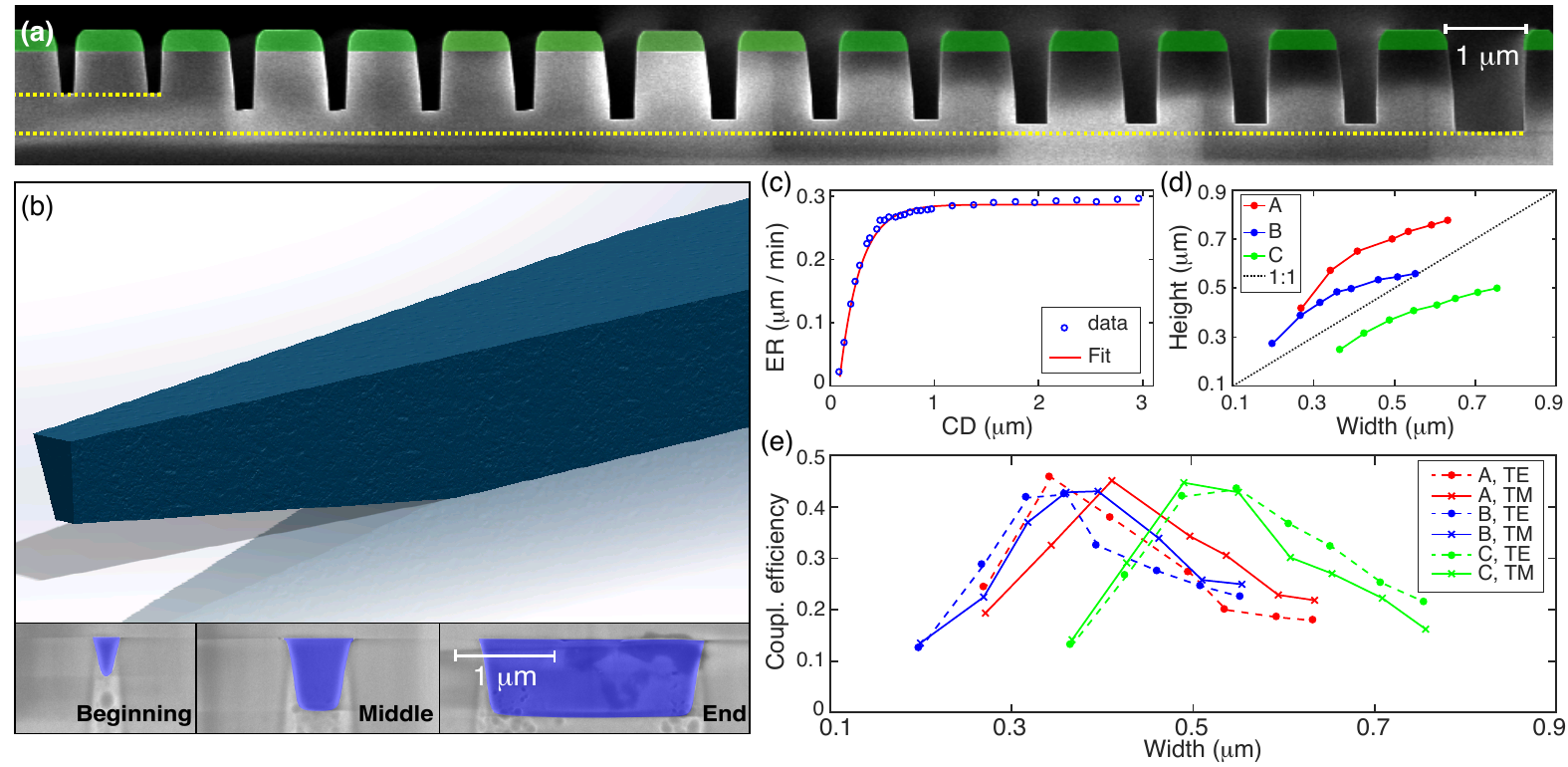}
\caption{Schematic and the characterization of the novel 2D double inverse tapers. (a) SEM image of the ARDE effect on a SiO$_2$ substrate with Si etch mask (green shaded). (b) Schematic of the 2D-taper. SEM images of the taper cross-sections (Si$_3$N$_4$ is blue shaded) at the taper beginning (chip facet side), middle, and end (bus waveguide side) are shown. (c) Characterized ARDE effect. SiO$_2$ is etched for 3 mins with a 400 nm Si etch mask. The blue circles are measured data via SEM and the red curve is the ARDE fit curve. CD: critical dimension of the trench width. ER: etch rate. (d) Measured height-to-width aspect ratios (HWAR) of 2D-tapers. A, B, C stand for 2D-taper chips 2D-A (red), 2D-B (blue) and 2D-C (green). (e) Measured coupling efficiency at 1550 nm wavelength (through two chip facets) of 2D-taper Chips A (red), B (blue), C (green), for both the TE (dash) and TM (solid) polarizations. By controlling the ARDE effect, the taper width at the optimum coupling can be shifted. Optimum coupling with >500 nm taper width can be achieved, enabling the use of normal UV lithography for the taper fabrication.}
\label{Fig:figure3}
\end{figure*}

In this Letter, we present novel, double inverse tapers ("2D-tapers") for efficient light coupling from lensed fibers to Si$_3$N$_4$ waveguides. In comparison with the regular inverse tapers ("1D-tapers") which have only the reduced waveguide width to match the incident mode (as shown in Fig. \ref{Fig:figure1}(a)), 2D-tapers have both the reduced waveguide height as well as the width (as shown in Fig. \ref{Fig:figure3}(b)). First, we present the simulation of light coupling from a lensed fiber to Si$_3$N$_4$ bus waveguides via 1D-tapers, using 3D finite-difference time-domain (FDTD) simulation. We fabricate 1D-tapers of 820 nm height but different widths, experimentally characterize their coupling efficiencies at 1550 nm wavelength, and demonstrate that >45$\%$ coupling efficiency for the transverse electric (TE) polarization can only be achieved in a taper of 80 nm width. Such small feature can only be achieved with electron beam lithography (EBL). We also reveal that the coupling efficiency for the transverse magnetic (TM) polarization is lower than the TE polarization due to the taper's large height-to-width aspect ratio (HWAR). To relax the stringent lithography requirement to fabricate tapers of 80 nm width and to further improve the coupling for the TM polarization, we fabricate 2D-tapers using the novel photonic Damascene process \citep{Pfeiffer:16}. Due to the inherent aspect-ratio-dependent etch (ARDE), the 2D-tapers have increasing height over taper length, therefore enabling further reduction of the taper size. We demonstrate >45$\%$ coupling efficiency at 1550 nm wavelength, for both the TE and TM polarizations in 2D-tapers of >300 nm width, which can be easily achieved with deep-UV lithography, significantly relaxing the lithography requirement. Furthermore, by comparing the coupling performance of several groups of 2D-tapers, we demonstrate the flexibility to engineer the 2D-taper's shape, enabling >45$\%$ coupling efficiency with $\sim$500 nm taper width. Therefore these 2D-tapers are promising for light coupling at short wavelength range e.g. 1064 nm or 780 nm, which is usually challenging as EBL is required to pattern very small taper size for optimized coupling. 

We experimentally study Si$_3$N$_4$ waveguides fully buried in SiO$_2$ cladding, as shown in Fig. \ref{Fig:figure1}(a). The 1D-tapers are fabricated using a subtractive process \cite{Gondarenko:09}, a standard CMOS-compatible fabrication process widely used for integrated photonic devices. In our process, patterns e.g. tapers and bus waveguides, are defined by EBL and transferred to Si$_3$N$_4$ film via dry etching (CHF$_3$/O$_2$). All the patterns, including tapers and bus waveguides, have the uniform height as the Si$_3$N$_4$ film thickness. As shown in Fig. \ref{Fig:figure1}(a), tapers fabricated using this process have increasing width but constant height over the taper length.

We simulate the light coupling from a lensed fiber to Si$_3$N$_4$ bus waveguides via 1D-tapers of different taper widths at a chip facet, using FDTD method. Due to the Si$_3$N$_4$ dry etching process, the 1D-tapers as well as the bus waveguides have a sidewall bottom angle of $\sim$80$^o$. In this Letter, the "taper width" is defined as the top side width of the taper's trapezoidal cross-section (see Fig. \ref{Fig:figure1}(a) inset), which is defined by EBL. Other taper parameters used in the simulation are shown in Fig. \ref{Fig:figure1}(c), (d). We use a free-space Gaussian mode of 2.5 $\mu$m waist diameter to represent the incident mode from the lensed fiber, according to the lensed fiber's specification. The normalized transmitted power through the bus waveguide, i.e. coupling efficiency per chip facet $T_1$, can be calculated in the simulation. Thus the full device coupling efficiency including two facets is defined as $T_2=T_1^2$.

We simulate $T_2$ at the wavelength 1550 nm, for both the TE and TM polarizations of the incident Gaussian mode. The simulated $T_2$ as function of taper width from 50 to 300 nm is shown in Fig. \ref{Fig:figure1}(b), together with the experimentally measured data which, will be discussed later.

The simulation results present two prominent trends: First, a smaller taper has better coupling, due to the weaker light confinement (lower effective refractive index) which improves the mode match to the incident Gaussian mode. Second, the TE mode has better coupling than the TM mode. As shown in Fig. \ref{Fig:figure1}(c), due to the taper's high height-to-width aspect ratio (HWAR), the TE mode has a larger size than the TM mode, leading to a better match to the incident Gaussian mode. Fig. \ref{Fig:figure1}(d) show the simulated mode propagation profile of the case of $\sim$80$\%$ coupling efficiency, illustrating that a small taper providing improved mode match can well guide the incident Gaussian mode to the bus waveguide.

We fabricate a large number of 1D-taper chips of $\sim$820 nm bus waveguide height (as the Si$_3$N$_4$ film thickness), and experimentally characterize their coupling efficiencies at 1550 nm wavelength, using a setup similar to the one described in Ref. \cite{pfeiffer:17}. The measured coupling efficiency of each sample, plotted in Fig. \ref{Fig:figure1}(b), agrees well with the simulated results, supporting the two aforementioned claims: A smaller taper has better coupling, and the TE mode has better coupling than the TM mode. The deviation between the measured data and the simulated results is likely due to the fact that the lensed fiber we use has non-unity transmission. We also characterize the coupling from 1500 to 1630 nm bandwidth, using a tunable laser. A weak trend of decreasing coupling efficiency with increasing wavelength is observed, as shown in Fig. \ref{Fig:figure1}(e). However this trend is more likely caused by the broadband response of e.g. the 50-50 fiber couplers used in the setup rather than the taper itself, as the opposite trend, increasing coupling efficiency with increasing wavelength, is observed when the 50-50 fiber coupler's two output branches are interchanged. Nevertheless, the coupling efficiency remains >30$\%$ over the 130 nm range. 




As mentioned above, 1D-tapers show polarization-dependent coupling, i.e. the TM polarization has lower coupling efficiency than the TE polarization, due to the 1D-taper's large HWAR. However, many photonic devices are specifically operated with TM mode, to couple vertically between different components, such as the coupling between photonic dielectric and plasmonic waveguides \cite{Christ:03, Briggs:10}. Therefore for these devices, efficient TM coupling is important and needs to be optimized.


Further improving the TM coupling requires reducing the taper size. However reducing the taper width further is challenging due to the lithography resolution and quality. In addition, thin but tall tapers of high HWAR tend to collapse which reduces fabrication yield. Therefore, reducing the taper height is a feasible solution, however simultaneously the height of other components, e.g. bus waveguides, should remain unchanged in order to operate the device in the same situation. As a result, tapers of both increasing width and height, manifesting as "double inverse tapers" or "2D-tapers", are desired. 

In the subtractive process all the patterns have the uniform height determined by the Si$_3$N$_4$ film thickness, thus 2D-tapers can not be achieved. Therefore, we use the photonic Damascene proess \citep{Pfeiffer:16}. In this process, the patterns are defined by lithography and then transferred to a SiO$_2$ substrate via dry etching. The Si$_3$N$_4$ film is then deposited on the SiO$_2$ substrate and fill the defined pattern trenches, followed by a chemical mechanical planarization (CMP) which removes the excess Si$_3$N$_4$ and planarizes the wafer top surface. As the dry etching process has inherently aspect-ratio-dependent etch (ARDE) rate \cite{Gottscho:92}, the pattern trench depth increases with increasing pattern size. For pattern sizes exceeding a certain threshold value, the ARDE effect becomes negligible, thus the trench depths can be considered as uniform. As a consequence, a taper of increasing width has increasing height, manifesting a 2D-taper. Other components, e.g. bus waveguides, have uniform height on the wafer.

\begin{table}[t!]
\vspace{-0.5cm}
\centering
\caption{\bf 2D-taper chips' specifications.}
\begin{tabular}{ccccc}
\hline
Chips & 1D & 2D-A & 2D-B & 2D-C\\
\hline
Mask thickness (nm) &  & $\sim$400 & $\sim$400 & $\sim$750\\
Bus waveguide height (nm) & $\sim$820  & $\sim$820 & $\sim$650 & $\sim$670\\
\hline
\end{tabular}
\label{tab:Taper}
\end{table} 

Fig. \ref{Fig:figure3}(a) shows the ARDE effect on a SiO$_2$ substrate, after SiO$_2$ dry etching process (C$_4$F$_8$/He) with a 400 nm amorphous Si etch mask. The ARDE effect creates non-uniform, pattern-dependent trench depths, and is characterized by measuring the mean etch rate (ER) as function of trench width (critical dimension, CD), as shown in Fig. \ref{Fig:figure3}(c). When CD reaches the threshold value of $\sim$1 $\mu$m, the ER reaches a stable value of $\sim$0.29 $\mu$m/min and becomes nearly independent of CD. From 0 to 0.5 $\mu$m CD, ER increases nearly linearly. The ARDE curve can be fitted with $\text{ER} = a\cdot \text{exp}(-\text{CD}/b)+c$, with $a=-0.444$ $\mu$m/min, $b=0.195$ $\mu$m, $c=0.286$ $\mu$m/min. Fig. \ref{Fig:figure3}(b) shows the measured SEM images of taper cross-sections at taper beginning (chip facet side), middle, and end (bus waveguide side), revealing the double inverse shape with increasing width and height of the waveguide.

Three groups of 2D-taper chips (Chips 2D-A, B, C) were fabricated and their specifications are shown in Tab. \ref{tab:Taper}, in comparison with the 1D-taper chip. These tapers' HWARs and coupling efficiencies for both the TE and TM polarizations are experimentally characterized, as shown in Fig. \ref{Fig:figure3}(d) and (e). Again, the taper width is defined as the taper's top side width. Compared with the 1D-taper chip shown in Fig.\ref{Fig:figure1}(b), Chip 2D-A achieves >45$\%$ TM coupling, while >45$\%$ TE coupling is maintained. Different from the 1D-tapers in which the smaller taper width shows better coupling, in Chip 2D-A, there is an optimum coupling point which is achieved with >300 nm taper width, due to the ARDE and the reverse trapezoidal shape (bottom angle 96$^o$). This optimum coupling requires weak waveguide confinement for large taper mode of improved mode match, but simultaneously the waveguide confinement needs to be sufficient to overcome the Gaussian beam's divergence. Such >300 nm taper width of optimum coupling can be easily achieved with deep-UV lithography instead of EBL, significantly relaxing the stringent requirement of lithography resolution. Moreover, optimized coupling for shorter wavelengths, e.g. 1064 nm and 780 nm, requires very small tapers, which is challenging to fabricate with 1D-tapers due to the waveguide height. Therefore 2D-tapers are more useful to work in these wavelengths. 

The ARDE and the taper shape can be engineered simply by changing the etch mask thickness. A thicker etch mask gives a stronger ARDE effect. To demonstrate this scheme, 2D-taper Chips, 2D-B and 2D-C, are fabricated. Chip 2D-C is fabricated with a 750 nm deep-UV photoresist etch mask, leading to a stronger ARDE. Chip 2D-B is fabricated with a 400 nm amorphous Si etch mask, same process as Chip 2D-A but the bus waveguide height is made to 650 nm via more CMP time, in order to directly compare to Chip 2D-C. Both chips have nearly the same bus waveguide height (660$\pm$10 nm), but their taper HWARs are significantly different due to their different ARDEs, as shown in Fig. \ref{Fig:figure3}(d). As shown in Fig. \ref{Fig:figure3}(e), compared with Chip 2D-A, the coupling of Chip 2D-B is not prominently different, because, as long as the taper's HWAR exceeds unity (>1), the mode size is more constrained by the taper width rather than the taper height. However, the stronger ARDE (HWAR <1) of Chip 2D-C shifts the optimum coupling to taper width >500 nm. In this case the required lithography precision is further reduced, enabling the use of common UV lithography.


In summary, we present the characterization of inverse (1D-) and double inverse (2D-) tapers for efficient light coupling from a lensed fiber to Si$_3$N$_4$ waveguides. We experimentally compare the coupling performance of 1D- and 2D-tapers, and illustrate the main mechanisms in the fabrication processes which lead to the performance difference. We demonstrate the advantages of 2D-tapers, including: First, better coupling of the TM polarization. Second, larger taper width at optimum coupling, enabling the use of UV or deep-UV photolithography instead of EBL. Third, flexibility to change the taper shapes via engineering the ARDE effect. The data presented in this Letter is from several chip devices, but is highly reproduceable in many other chips that we have experimentally characterized. Our results demonstrate the advantages of 2D-tapers over 1D-tapers, particularly promising for light coupling at near-IR or visible wavelengths.  

\subsection*{Funding Information}
Defense Advanced Research Projects Agency (DARPA), Defense Sciences Office (DSO) (HR0011-15-C-0055); European Union\textquoteright s Horizon 2020 Framework Programme (H2020) (709249); Swiss National Science Foundation (SNSF) (161573). 

\subsection*{Acknowledgements}
We thank Tiago Morias in assisting the sample fabrication. The Si$_3$N$_4$ optical waveguide samples were fabricated in the EPFL center of MicroNanoTechnology (CMi).

\bibliographystyle{apsrev4-1}
\bibliography{Bibcolec}
\end{document}